\documentclass[twocolumn]{article} 

\usepackage[superscript,biblabel]{cite}

\usepackage{caption}
\usepackage{sectsty}
\chapterfont{\raggedright\normalsize\MakeUppercase}
\sectionfont{\raggedright\normalsize\MakeUppercase}
\subsectionfont{\raggedright\large}
\subsectionfont{\raggedright\normalsize}

\usepackage[toc,page]{appendix}

\usepackage[left=2.6cm,right=2.3 cm,top=1.8cm,bottom=3.6cm]{geometry}
\usepackage{amsmath}
\usepackage{amssymb}

\usepackage{hyperref}
\hypersetup{colorlinks = true, linkcolor = blue, citecolor = blue, urlcolor = black}

\usepackage{graphicx}
\usepackage{xcolor}

\newcommand{\be}{\begin{equation}}
\newcommand{\ee}{\end{equation}}
\newcommand{\bea}{\begin{eqnarray}}
\newcommand{\eea}{\end{eqnarray}}

\newcommand{\units}[1]{{~\rm #1}}

\newcommand{\fig}[1]{Fig.\,\ref{#1}}
\newcommand{\Fig}[1]{Figure\,\ref{#1}}

\newcommand{\eq}[1]{Eq.\,(\ref{#1})}
\newcommand{\Eq}[1]{Equation (\ref{#1})}

\newcommand{\B}{{\bf B}}

\renewcommand{\r}{{\bf r}}
\renewcommand{\k}{{\bf k}}
\renewcommand{\d}{\mathrm{d}}

\newcommand{\lc}{l_c}
\newcommand{\tc}{t_c}
\newcommand{\Om}{\overline\Omega}

\usepackage{soul}

\usepackage{authblk} 

\title{Larmor Frequency Depends on Structural Anisotropy in Magnetically Heterogeneous Media}

\author[1,2]{Alexander Ruh}
\author[1]{Valerij G. Kiselev}
\affil[1]{Medical Physics, Dept.\ of Radiology, Medical Center -- University of Freiburg, Faculty of Medicine, University of Freiburg, Germany}
\affil[2]{Dept. of Radiology, Feinberg School of Medicine, Northwestern University, Chicago, IL, United States}

\begin{document}

\twocolumn[
\begin{@twocolumnfalse}

\maketitle

\begin{abstract}
\noindent
{\bf Purpose:} To investigate the effect of anisotropic magnetic microstructure on the measurable Larmor frequency offset in media with heterogeneous magnetic susceptibility. Specific objectives were (i) validation of recently developed theory for the case of fast diffusion and (ii) investigation of the transition between the regimes of fast and slow diffusion. \\
{\bf Methods:} Monte Carlo simulations in synthetic media. \\
{\bf Results:} Simulations demonstrate a perfect agreement with the previously developed theory for fast diffusion. Beyond this regime, the frequency offset shows a pronounced dependence on the medium microarchitecture and the diffusivity of NMR-reporting spins in relation to the magnitude of the susceptibility-induced magnetic field. \\
{\bf Conclusion:} While the effect of myelin in brain white matter is commonly treated assuming efficient diffusion narrowing, this regime does not hold for larger cells or higher magnetic susceptibility. In such a case, the effect essentially deviates from the prediction based on the assumption of diffusion narrowing. 
\bigskip 

\end{abstract}


\end{@twocolumnfalse}\vspace{1cm}

]


\section{Introduction}
\label{sec:Intro}

Precise signal phase measurements \cite{Abduljalil2003,Duyn2007,Marques2009}  initiated a still ongoing discussion of the microstructural correlates of the proton precession frequency in brain white matter \cite{He2009,Yablonskiy2012,Sukstanskii2014,Duyn2014_wm,Yablonskiy2014_wm,Duyn2014,Yablonskiy2014,Yablonskiy2015,Yablonskiy2017,Duyn2017,Duyn2018,Yablonskiy2018}. In a broader view, the question is about the averaged Larmor frequency of an ensemble of spins, moving in a magnetically heterogeneous medium, for instance, containing microscopic inclusions with a distinct magnetic susceptibility (e.g.\ biological cells). While the case of fast moving spins, the so-called {\em diffusion narrowing regime} (DNR), remains in the discussion focus, a few studies addressed the opposite regime of negligible diffusion, the so-called {\em static dephasing regime} (SDR) \cite{Yablonskiy94,Kokeny2018,Ruh2018}. To the best of our knowledge there is a single study of the transition between these limiting cases for the case of isotropic media \cite{Ruh2018}. 

In the present study this transition is investigated in silico for anisotropic media. Our simulations demonstrate the importance of the {\em structural anisotropy} of the medium microarchitecture. For a given medium microstructure, the transition is governed by a combined parameter describing the typical phase accumulation by a spin diffusing over the characteristic microstructural length scale. The simulations also validate the recently developed theory for anisotropic media in the DNR \cite{Kiselev2018}.

\section{Theory}
\label{sec:Theory}

We consider media with impermeable, NMR-invisible inclusions with the isotropic magnetic susceptibility $\chi$ relative to the surrounding NMR-visible fluid (Fig.1). Exposed to the external field $\B_0$ the inclusions create a local Larmor frequency shift $\Omega(\r)$, which is the most straightforward to calculate in terms of the Fourier transformed quantities, 
\begin{equation} \label{Om=}
\Omega(\k)=\delta\Omega\,Y(\k)\,v(\k)\,,\quad\delta\Omega=4\pi\gamma |\B_0|\chi\,,
\end{equation}
where $\gamma$ is the proton gyromagnetic ratio, and $v(\k)$ is the Fourier transform of $v(\r)$, the indicator function that is unity inside the susceptibility inclusions and zero otherwise. Throughout this paper we use the same letters for real-space and Fourier-trans\-formed quantities; the argument is always given explicitly to avoid confusion. The magnitude of the Larmor frequency fluctuations, $\delta\Omega$, is written in the cgs system; in SI, the factor $4\pi$ is absorbed in the correspondingly larger value of $\chi_{\rm SI} = 4\pi \chi$. The function $Y(\k)$ is proportional to the elementary dipole field, 
\be  \label{defY}
Y(\k)=\frac{1}{3}-\frac{k_3^2}{k^2}\,, 
\ee
where the third direction is selected along the main magnetic field. In what follows, the local Larmor frequency, $\Omega$, is referred to as {\em the field}.

\begin{figure}
\includegraphics[width=\columnwidth]{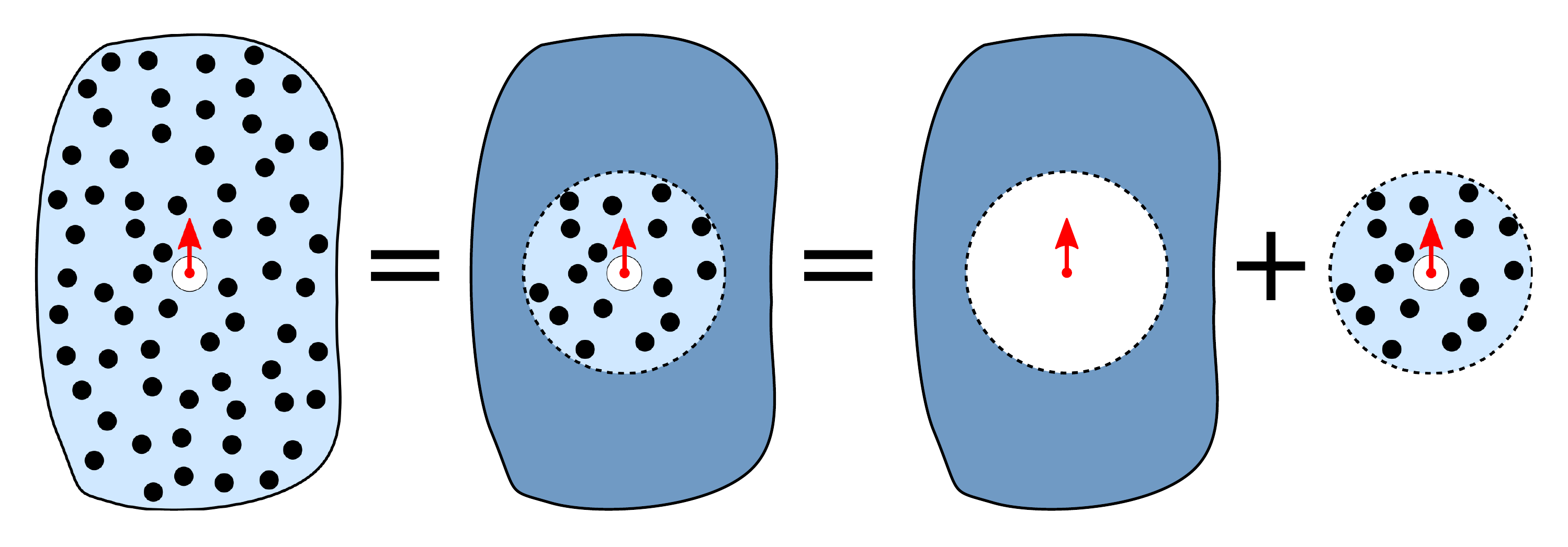}
\caption{\label{fig:sample_decomposition}
A sample with a microscopic magnetic structure and the decomposition of its magnetization for calculating the magnetic field at the position of an NMR-reporting spin (red arrow). The heterogeneous magnetization is shown as inclusions (black dots) with a magnetic susceptibility, $\chi$, different from that of the surrounding fluid (light blue). The first equality: The sample can be divided into a far and a spherical near region, the {\em mesoscopic sphere} (the sphere size is greatly exaggerated in the image). The far region contributes a field induced to its average (homogenized) susceptibility (darker blue). The second equality: The field at the spin's position is the sum of two contributions induced by the far region with excluded mesoscopic sphere and the mesoscopic sphere. While the former is governed by the macroscopic averaged magnetic susceptibility, the latter depends on the medium microarchitecture. The small empty sphere stands for the classical Lorentz sphere in the NMR-reporting fluid.
}
\end{figure}

Since the field at any point in a sample depends on the overall sample shape, our analysis is applied to a small volume of the standard spherical form drawn around a given spin. This volume is very small on the macroscopic scale of imaging, but contains a statistically large number of susceptibility inclusions; we refer to this volume as {\em the mesoscopic sphere} \cite{Ruh2018}. In real experiments, the field created by the rest of the sample should be added as illustrated in \fig{fig:sample_decomposition}. 

Let the frequency offset derived from the signal phase within the mesoscopic sphere be $\Om$. The way how the heterogeneous field, $\Omega(\r)$, is averaged in the signal giving rise to $\Om$ depends on diffusion in the NMR-visible fluid. When diffusion is fast, the typical spin samples the spatially averaged field, $\left< \Omega(\r) \right>$, and the normalized signal is just $S = \exp\left[ -i\left< \Omega(\r) \right> t\right]$. In the opposite case of static dephasing, the signal is the mean of individual spins' phase factors, $S = \left< \exp \left[ -i\Omega(\r) t\right]  \right>$.

The transition between the two limiting cases is governed by a single parameter, \cite{Ruh2018}
\begin{equation}\label{def_varphi}
\varphi = {\delta\Omega \, \lc^2\over D} \, ,
\end{equation}
where $D$ is the isotropic diffusivity in the fluid and $\lc$ the microstructural correlation length in the medium;  for dense packings of compact objects, this length is defined by their size. The controlling parameter $\varphi$ is the typical phase acquired by a spin diffusing past a single susceptibility inclusion. For example, the DNR condition $\varphi\ll 1$ can be fulfilled due to a weak field, $\Omega(\r)$, or due to fast diffusion that can efficiently average a stronger field. 

The distinct role of $\varphi$ follows from the Bloch-Torrey equation \cite{Torrey56} for the complex-valued transverse magnetization $\psi$. This equation can be cast in a dimensionless form in terms of variables $\xi=\r/\lc$, and $\tau=t/\tc$, where $\tc=\lc^2/D$ is the typical time for spins to diffuse across field variations of size $\lc$: 
\begin{equation}\label{BT_dl}
\frac{\partial}{\partial \tau} \, \psi = \nabla_\xi^2 \, \psi - i \, \frac{\delta \Omega \, l_c^2}{D} \, \frac{\Omega(\r)}{\delta \Omega} \, \psi \,.
\end{equation} 
The magnitude of the normalized field $\Omega(\r)/\delta \Omega$ is independent of $\chi$. This field embodies the statistics of the original medium scaled in such a way that the field variations occur over the typical distance $\xi\sim 1$. Note that rescaling this length must be accompanied by a proportional rescaling of all medium dimensions. For the standard initial condition, $\psi|_{\tau=0}\equiv1$, three medium parameters enter the solution via the single dimensionless combination, \eq{def_varphi}. This fact is referred to as {\em scaling}.  

\begin{figure}
\includegraphics[width=\columnwidth]{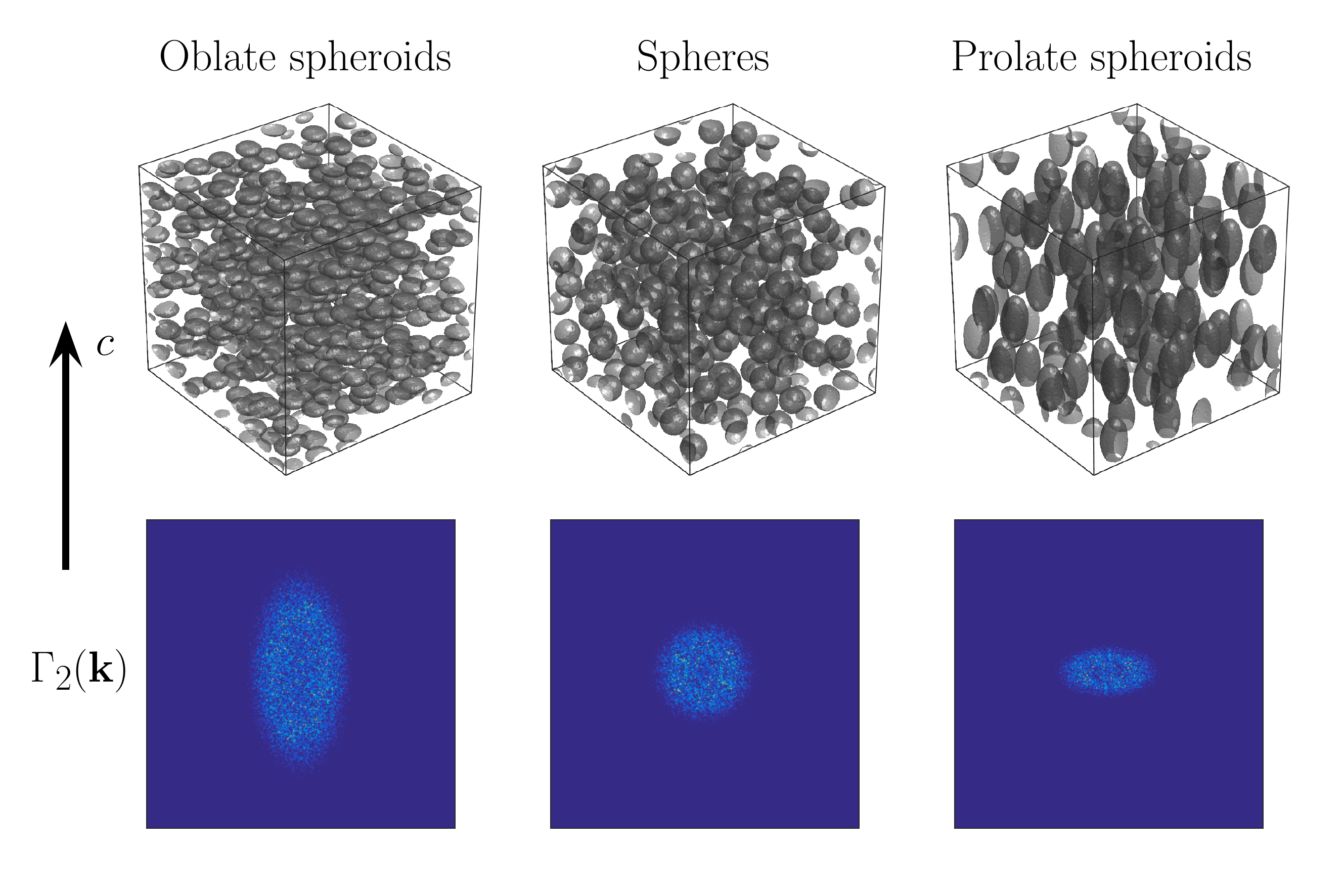}
\caption{\label{fig:3media}
Top row: Examples of simulated media consisting of identical magnetic susceptibility inclusions shaped as oblate spheroids  (axes ratio $c/a=0.5$), spheres and prolate spheroids ($c/a=2$); $a$ is the same for all three. The visualization shows a quarter of the simulation box in each dimension. Bottom row: The corresponding Fourier-transformed correlation functions $\Gamma_2(\k)$. The visualized cross-sections are parallel to the symmetry axis and zoomed 4-fold around the origin.
}
\end{figure}

Consider the diffusion narrowing regime. The mean Larmor frequency can be calculated analytically by performing the averaging over the space outside the susceptibility inclusions with the result \cite{Ruh2018,Kiselev2018}  
\begin{equation} \label{theoryDNR}
\Om = \langle \Omega \rangle = -\frac{\delta\Omega}{1-\zeta} \int \! \frac{\d^3 k}{(2\pi)^3} \; \Gamma_2(\k) \, Y(\k) \, ,
\end{equation}
where $\zeta$ is the volume fraction of the susceptibility inclusions with $\zeta=\left< v(\r)\right>$, and $\Gamma_2(\k)$ is the Fourier transform of the two-point correlation function of inclusions, 
\begin{equation}\label{defGamma}
\begin{split}
& \Gamma_2(\r) = \langle v(\r_0 + \r) \, v(\r_0) \rangle_{\r_0}-\zeta^2 \,,  \\ 
& \Gamma_2(\k \neq 0) = \frac1V \, v(\k) \, v(-\k) \,, \quad \Gamma_2(\k = 0) = 0\,.
\end{split}
\end{equation}
This correlation function was considered in detail in Refs.\,\citen{Novikov2008,Burcaw2015} and is illustrated in \fig{fig:3media}.

\Eq{theoryDNR} describes the effect of structural anisotropy in a medium built with isotropic materials (both the fluid and the inclusions). For the case of isotropic structure, $ \Gamma_2(\k)$ is spherically symmetric (\fig{fig:3media}) and the integral in \eq{theoryDNR} turns to zero due to the form of $Y(\k)$, \eq{defY}. Otherwise, \eq{theoryDNR} predicts a finite frequency shift. It can be obtained in general terms for the case of cylindrical geometry, that is for the structure independent of one spatial direction, $c$, with the axial symmetry of $\Gamma_2(\r)$ in the transverse plane \cite{Kiselev2018}. In brief, $\Gamma_2(\k)$ in this case is proportional to $\delta(k_c)$, which helps to perform the integration in \eq{theoryDNR}. Since $Y(\k)$, \eq{defY}, is proportional to the spherical harmonics $Y_2^0$, the only relevant components of the $\Gamma_2(\k)$ orientation dependence are those with the angular momentum $\ell=2$, from which the axial symmetry leaves only one with the projection $m=0$. However, this applies to the spherical harmonics defined relative to the $c$-axis, while the dipole field, $Y(\k)$, follows the direction of the main magnetic field. Performing the rotation of spherical harmonics to the same axis gives \cite{Kiselev2018}
\be \label{meso_cyl}
\overline \Omega = -\frac{\zeta}{2} \left(\cos^2\theta - \frac 13\right) \delta\Omega\,,
\ee
where $\theta$ is the angle between the main magnetic field and the symmetry axis of the medium. Detailed analysis and generalization for anisotropic materials are presented in Ref.\,\citen{Kiselev2018}.

Less is known about the frequency shift in the static dephasing regime. Analytical expressions are available for randomly placed spheres or cylinder with an overall low volume fraction, $\zeta\ll 1$ \cite{Yablonskiy94}. The result for spheres is accurate for $\zeta \lesssim 0.2$ \cite{Ruh2018}.

\section{Methods}
\label{sec:Methods}

Disordered three-dimensional media consisting of non-overlapping, identical inclusions with a distinct magnetic susceptibility were generated using random sequential addition with periodic boundary conditions in all three spatial directions (\fig{fig:3media}). The volume fraction of inclusions was $\zeta=0.15$. The inclusions had the form of spheroids with the semiaxes $c\,,a\,,a$ with a given aspect ratio $c/a$. Different media were generated for $c/a$ selected as powers of 2 in the range from $1/8$ to $16$. To obtain structural anisotropy, the orientation of the principal axis, $c$, was restricted to a narrow cone with solid angle $0.008\units{sr}$. 

The generated media were sampled on a $1024^3$ cubic grid for numerical computations of the field and successive Monte Carlo (MC) simulations. The inclusion size was chosen from $a = 7 \Delta x$ for prolate to $a = 28 \Delta x$ for oblate spheroids to keep their volume comparable, where $\Delta x$ was the grid spacing. The local precession frequency was calculated according to a discretized version of \eq{Om=} with the $\k=0$ component explicitly set to zero \cite{Ruh2018}. For media consisting of oriented spheroids, the magnetic field was generated twice, parallel and perpendicular to the object orientation (the $c$ axis).

Diffusion was simulated by random hopping on the grid with random walkers initialized outside the inclusions and kept there by prohibiting the penetration inside the inclusions. The complex-valued free induction decay signal was calculated as the mean of accumulated phase factors of individual random walkers at every time point. The number of random walkers was $10^6$  for fast diffusion and increased to $10^7$ for $\varphi>100$, \eq{def_varphi}, to compensate for less diffusion averaging in this regime. The frequency shift, $\Om$, was determined by the maximum of the spectral line obtained from the Fourier transformation of the signal time course\cite{Ruh2018}.

\begin{figure}
\centering
\includegraphics[width=0.8\columnwidth]{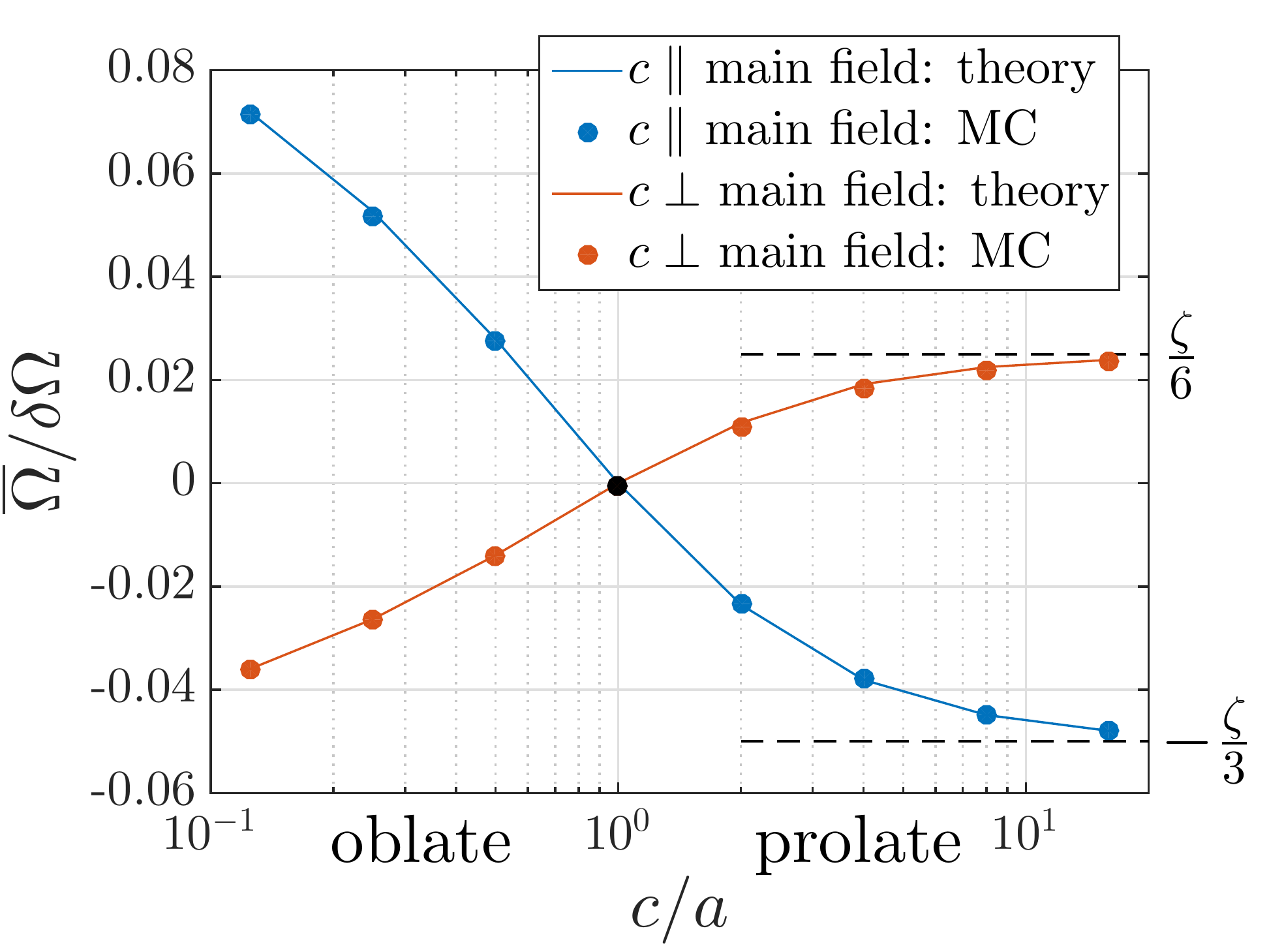}
\caption{\label{fig:Omega}
The frequency shift, $\overline\Omega/\delta\Omega$, for fast diffusion ($\delta\Omega\,l_c^2/D\lesssim4$) for different media, characterized by their aspect ratio $c/a$ (abscissa) for parallel (blue) and perpendicular (red) orientations to the main field. Solid lines show the results of numerical integration of the correlation function according to \eq{theoryDNR}. The prediction does not involve any parameter fitting. Circles show results of Monte Carlo simulations in the same media demonstrating an excellent agreement with theory. Note the black circle showing zero frequency shift for isotropic (spherical) inclusions ($c/a=1$). When the prolate spheroids get longer, the frequency shift approaches the theoretically calculated  limits of infinitely long cylinders, $\theta=0$ and $\theta=\pi/2$ in \eq{meso_cyl} (dashed lines).
}
\end{figure}

\begin{figure*}
\centering
\includegraphics[width=\textwidth]{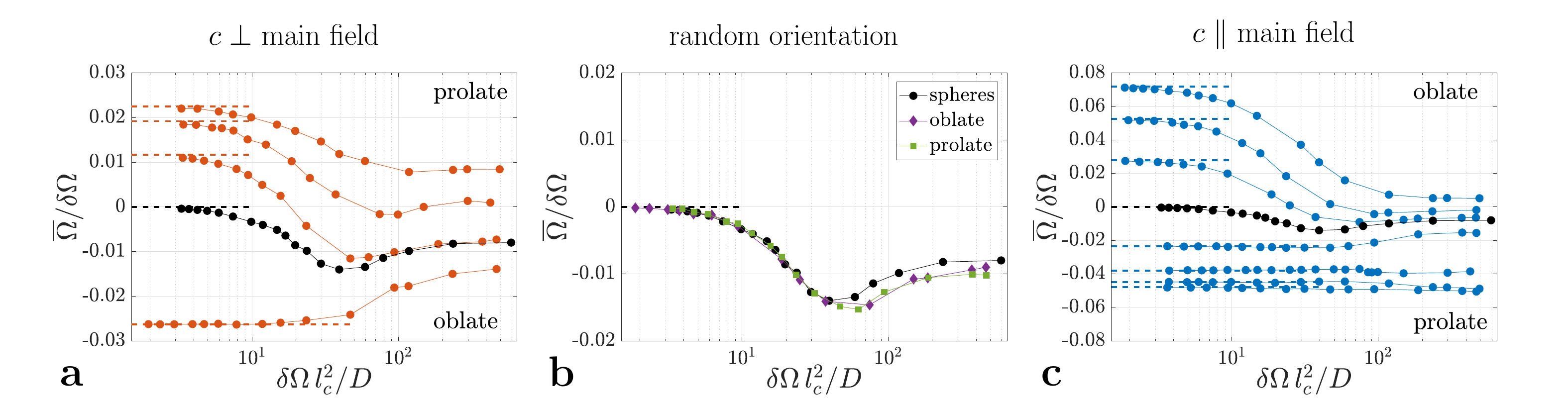}
\caption{\label{fig:transition}
The frequency offset shows a pronounced dependence on the regime-controlling parameter $\varphi = \delta\Omega \lc^2/D$, \eq{def_varphi} (abscissas) and the inclusions's shape and orientation. For small values of $\varphi$ all results approach the theoretical DNR limit, \eq{theoryDNR} (dashed lines). Results for spherical inclusions are reproduced in black in all panels. \textbf{a}: Inclusions with coherent orientations with the symmetry axis $c$ perpendicular to the main field. The inclusions' shape is changed from prolate to oblate spheroids (top to bottom). \textbf{b}: Inclusions with $c/a = 0.5$ and 2 with random orientations. \textbf{c}: The media from panel (\textbf{a}) plus three more for the field parallel to the symmetry axis (all media from \fig{fig:Omega}). The result for the most prolate inclusions approaches the value for long cylinders in the mesoscopic sphere, $-\zeta/3=-0.05$. 
}
\end{figure*}

\section{Results}
\label{sec:Results}

We first investigated the microstructure dependence of the frequency shift for fast diffusion (in the DNR). \Fig{fig:Omega} shows the result of Monte Carlo simulations for various aspect ratios of spheroids. Comparison with the deterministic theory expressed by \eq{theoryDNR} demonstrates a perfect agreement without any parameter fitting. 

The transition from the DNR to the SDR is illustrated in \fig{fig:transition}. The result shows a pronounced dependence on the medium structure and the value of the controlling parameter $\varphi$, \eq{def_varphi}. The results for spheroids with random orientation are very close to that for spheres. For the DNR, this follows from \eq{theoryDNR}, which is not sensitive to the origin of an isotropic correlation function, either due to the spherical inclusions or randomized non-spherical ones. A light deviation from this equivalence is observed in the SDR. 

The $\varphi$-dependence flattens out when the medium approaches the case of long cylinders parallel to the main magnetic field. The limiting $\varphi$-independent value is $-\zeta/3$ according to \eq{meso_cyl} with $\theta=0$. This corresponds to the well-known fact that a bunch of cylinders parallel to the main field does not create any field in the space between cylinders. In such a sample, the volume external to the mesoscopic sphere (the first term on the right-hand size of the graphical equation shown in \fig{fig:sample_decomposition}) has a cylindrical shape with a void, which results in the precisely opposite contribution \cite{Kiselev2018}. Since this contribution is governed by the averaged magnetic susceptibility, it is insensitive to the microstructure and so is the compensating field of the mesoscopic sphere.

\section{Discussion} 
\label{sec:Discussion}

This study provides a validation for the recently obtained analytical expression, \eq{theoryDNR}, describing the frequency shift in the diffusion narrowing regime. Dependence of the precession frequency on the microstructure in this regime was first discussed in the context of NMR in anisotropic solutions and liquid crystals \cite{Buckingham60,Haller73,Palffy-Muhoray77,Dunmur83}. A pronounced precession frequency dependence on the microstructure was obtained using MC simulations for the model of multiple sclerosis in which the same amount of magnetic material was redistributed from hollow cylinders representing the intact myelin sheets to isotropically distributed debris \cite{Yablonskiy2012}. The dependence on microstructure was confirmed in the following theoretical studies \cite{Sukstanskii2014,Yablonskiy2015,Duyn2017} and discussed in review series \cite{Yablonskiy2017,Duyn2017,Duyn2018,Yablonskiy2018}. 

In more detail, the present study deals with structurally anisotropic media in which the anisotropy is achieved exclusively by the form of inclusions, while the magnetic susceptibility of inclusions' material is isotropic. According to \eq{theoryDNR}, possible anisotropy in the arrangement of susceptibility inclusions enters via a single quantity, the inclusions' correlation function, which represents the medium structure in a unified way. We did not consider the effect of molecular-level anisotropy described by the magnetic susceptibility tensor of the inclusions' material \cite{Liu2010,Lee2010,Wharton2012,Sukstanskii2014,Yablonskiy2017,Duyn2017}; a generalization of \eq{theoryDNR} for this case was obtained recently \cite{Kiselev2018}. 

The precession frequency in the opposite limiting case of the static dephasing also shows dependence on the microstructure \cite{Yablonskiy94,Chen2013}. Less is known about the transition between the DNR and the SDR. It was studied using MC simulations for isotropic media built with spherical inclusions and experiments with microbead suspensions \cite{Ruh2018}. The present study extends the previous results for anisotropic media demonstrating a similar pattern of the transition. 

Considering biological tissues, the present results are relevant for the extracellular compartment. The intracellular and, possibly, myelin signal should be added to describe the whole tissue \cite{Sukstanskii2014,Yablonskiy2017,Duyn2017,Duyn2018,Yablonskiy2018}. The diffusion narrowing regime is a good approximation for the native magnetic susceptibility of myelin with the axonal sizes of the order of a micrometer \cite{Sukstanskii2014,Duyn2017,Ruh2018}. According the the scaling expressed by the controlling parameter $\varphi = \delta\Omega \lc^2/D$, the signal phase accumulation measured in brain white matter cannot be straightforwardly translated to tissues with larger axons or enhanced magnetic susceptibility of cells or reduced diffusion coefficient in ex-vivo samples. The only exception from this warning is the case of axons parallel to the main field although the effect of axonal orientation dispersion \cite{Ronen2014} remains to be studied.

\section*{Acknowledgement}
This work was supported by the German Research Foundation (DFG), grant KI\,1089/6-1.

\bibliographystyle{vancouverAR}
\bibliography{c:/Users/ruhal/Dropbox/Work/Literatur/bibtex/library}

\end{document}